\documentclass[10pt]{IEEEtran}

\usepackage{amsmath}
\usepackage{mathtools}
\usepackage{graphicx}
\usepackage{enumerate}
\usepackage{url}
\usepackage{subfigure}
\usepackage{algorithm}
\usepackage{algpseudocode}
\usepackage{stfloats}
\usepackage{lineno}
\usepackage{hyperref}
\usepackage{bm}
\usepackage{bbm}
\usepackage{amssymb}
\usepackage{enumerate}
\usepackage{xcolor}
\usepackage{float}
\usepackage{cite}
\usepackage{tabularx}
\usepackage{tabu}
\usepackage{multicol}
\usepackage{multirow}
\usepackage{colortbl,booktabs,threeparttable}
\usepackage{dcolumn}
\usepackage{flushend}
\usepackage{soul}
\usepackage{ragged2e}

\graphicspath{{Figures/}}

\definecolor{hl-bg-color}{RGB}{255,255,215}
\sethlcolor{hl-bg-color} 
\definecolor{new-magenta}{RGB}{255,0,255}
\soulregister\cite7
\soulregister\citep7
\soulregister\citet7
\soulregister\ref7
\soulregister\eqref7
\newcommand*{\HIGHLIGHT}{}
\ifdefined\HIGHLIGHT

\else

\fi

\newcommand{\thickhline}{%
    \noalign{\hrule height 0.8pt}
}

\newcommand*{\IEEE}{}
\ifdefined\IEEE

\fi
\begin{document}
\newpage
\title{Resilience in Trustworthy Wireless Systems}

\author{Shixiong Wang,~
        Yumeng Zhang,
        and Hongyu Li
\thanks{
S. Wang is with the School of Mathematics and Statistics, Xi'an Jiaotong University, Xi'an 710049, China (E-mail: s.wang@xjtu.edu.cn; s.wang@u.nus.edu).
Y. Zhang is with the Department of Electronic and Computer Engineering, Hong Kong University of Science and Technology, Hong Kong (E-mail: eeyzhang@ust.hk).
H. Li is with the Internet of Things Thrust, Hong Kong University of Science and Technology (Guangzhou), Guangzhou, China (E-mail: hongyuli@hkust-gz.edu.cn).
(\textit{S. Wang and Y. Zhang contribute equally; Corresponding Author: H. Li.})
}
}

\maketitle

\begin{abstract}
Resilience has emerged as a fundamental capability for future wireless systems operating in dynamic and uncertain environments. Although resilience has attracted growing attention across academia, industry, and standardization, its conceptual scope, enabling mechanisms, and realization techniques remain  fragmented. This paper presents a systematic framework of resilience in wireless systems from a trustworthiness perspective. We first formalize the concept of resilience and distinguish it from related uncertainty-aware terminologies, including reliability, robustness, adaptability, survivability, and recoverability. We then establish a hierarchical framework that organizes resilience into capability dimensions and enabling mechanisms, and that quantifies resilience through different technical aspects. 
We further use physical links and unmanned aerial vehicle networks as representative wireless scenarios to demonstrate how resilience can be systematically realized through the joint design of architectures, operations, and algorithms. Finally, we examine the fundamental trade-offs in resilience engineering: Improving resilience generally incurs costs in resource efficiency, nominal performance, information acquisition, implementation complexity, and latency.
\end{abstract}


\section{Introduction}\label{sec:introdction}
\begin{figure}[!htbp]
    \centering
    \includegraphics[width=0.48\textwidth]{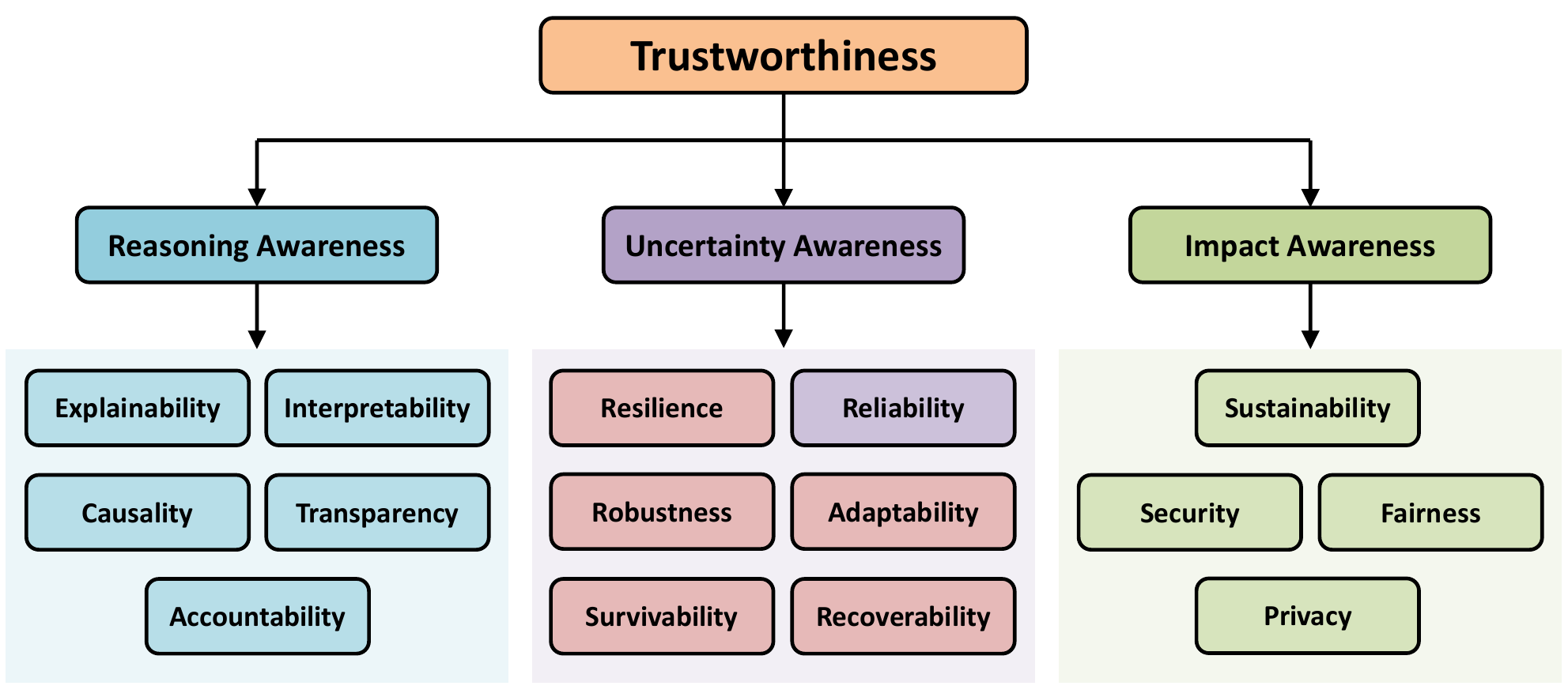}
    \caption{Overarching aspects of trustworthy wireless systems.}
    \label{fig:trustworthiness}
\end{figure}

Future wireless systems, such as cellular, sensor, industrial ad-hoc, and internet-of-things (IoT) networks, are anticipated to possess diverse capabilities that are reasoning-aware of the inference mechanisms behind their decisions, uncertainty-aware of dynamic and uncertain operating environments, and impact-aware of the environmental and societal implications of their deployment. To be specific, reasoning awareness primarily encompasses explainability, interpretability, transparency, causality, and accountability \cite{guo2020explainable,thomas2024causal,xiao2008accountability}. Uncertainty awareness mainly entails resilience, reliability, robustness, adaptability, survivability, and recoverability \cite{wang2025uncertainty,wang2025fast,ma2026vision}. Impact awareness mostly involves sustainability (e.g., greenness), privacy, security, and fairness \cite{rahmani2023next,chen2011fundamental,qu2018privacy,zou2016survey,huaizhou2013fairness}. Collectively, these complementary capabilities constitute the blueprint of trustworthy wireless systems; see Fig. \ref{fig:trustworthiness}. This article focuses on the principles and methodologies of resilience in wireless systems.

In November 2023, the International Mobile Telecommunications towards 2030 and beyond (IMT-2030) identified ``resilience" as one of the fundamental capabilities for future networks, where ``resilience refers to the capabilities of networks and systems to continue operating correctly during and after a natural or man-made disturbance, such as the loss of a primary source of power" \cite[p.~16]{IMT-2030}. 
To be specific, in the newly introduced IMT-2030 usage scenario of ``ubiquitous connectivity'', supporting systems are required to be resilient to, e.g., external disturbances and internal faults. In addition to resilience, closely related terminologies with respect to uncertainty awareness, as frequently discussed in academia (e.g., \cite{wang2025uncertainty}) and industry (e.g., \cite{ma2026vision}), also include reliability, robustness, adaptability, survivability, recoverability, etc. However, the conceptual boundaries and relationships among these terminologies remain fragmented and lack a unified study. Moreover, despite growing research interest since the IMT-2030 initiatives, the principles and methods of resilience in wireless systems are to be systematically investigated.

To bridge the two gaps above, this overview article explores the basic connotations, enabling mechanisms, and concrete techniques of resilience in wireless systems, and establishes a unified perspective on its relationships with other intrinsically coupled concepts under the uncertainty-awareness paradigm.

\section{Fundamentals of Resilience}\label{sec:concepts}
\begin{figure}[!htbp]
    \centering
    \includegraphics[height=4cm]{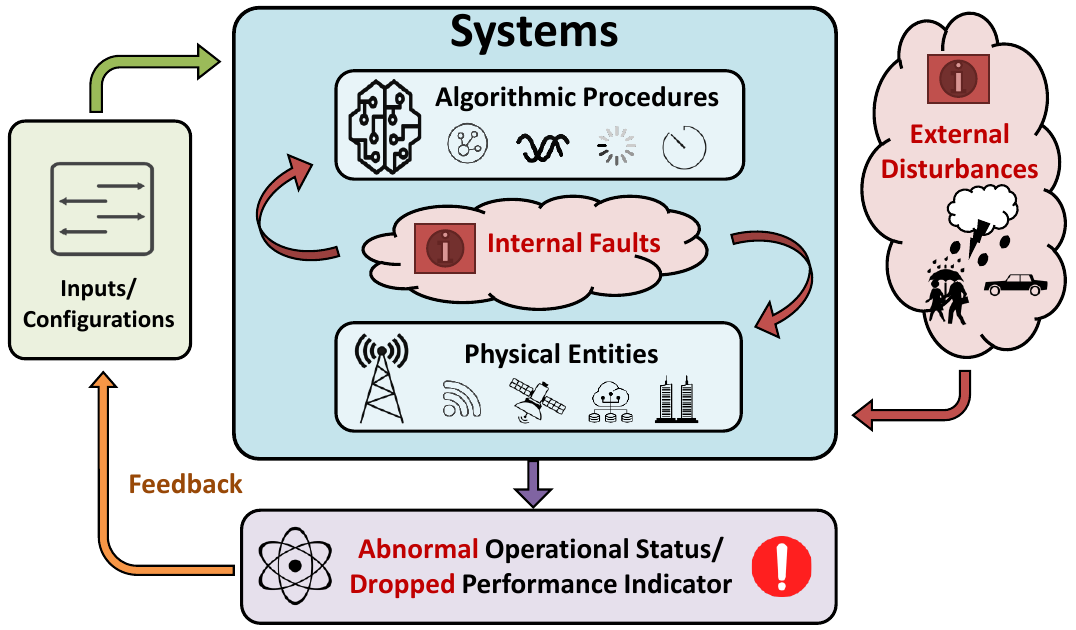}

    \caption{The building diagram of a system subject to uncertainties, including both external disturbances and internal faults. Uncertainties lead to abnormal operations and dropped performance.}
    \label{fig:building_diagram}
\end{figure}

\begin{figure}[!htbp]
    \centering
    \includegraphics[height=5.5cm]{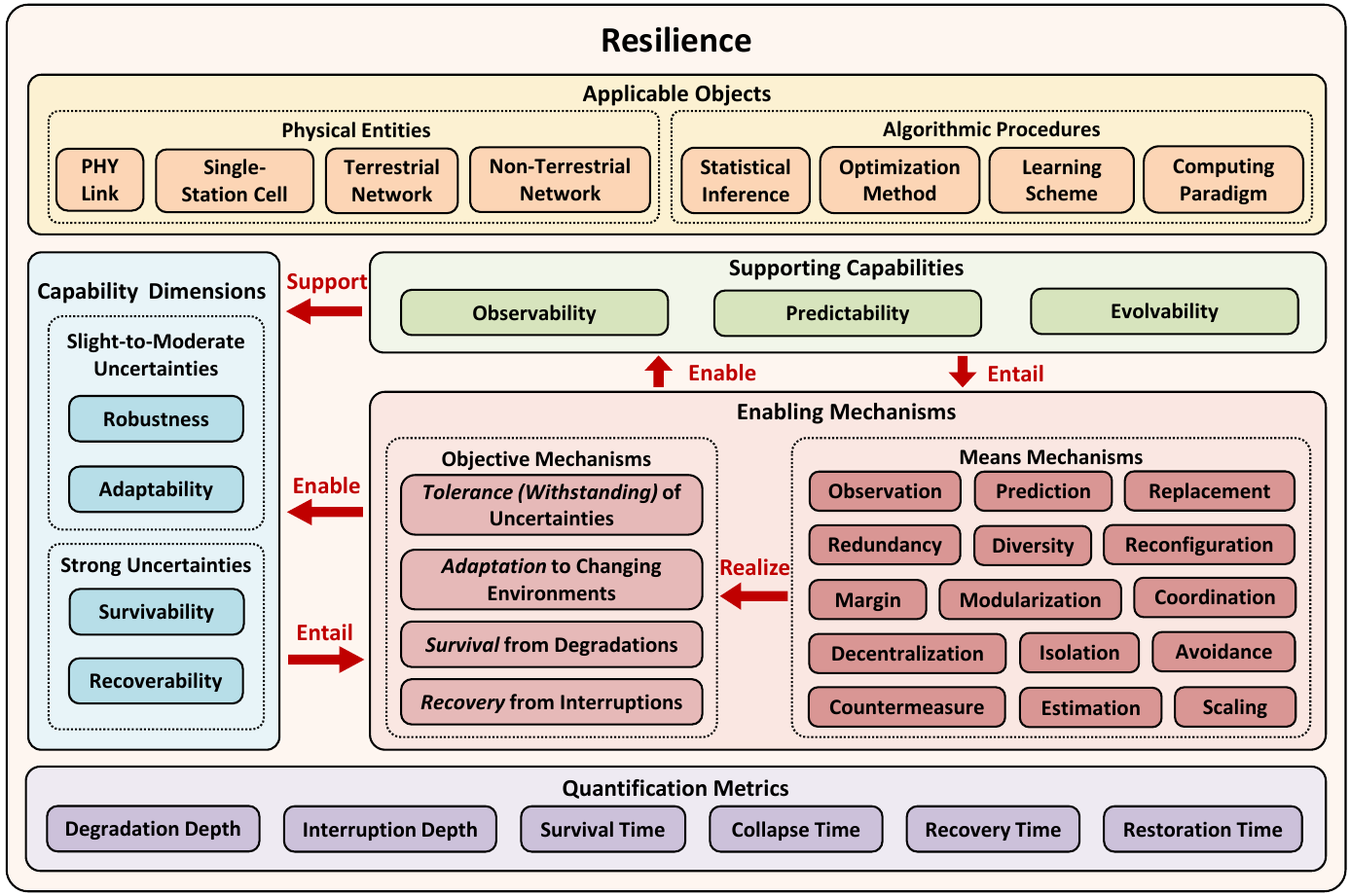}
    \caption{Fundamentals of resilience, including its applicable objects, relationships to other terminologies, enabling mechanisms, and quantification metrics.}
    \label{fig:key_information_fundamentals}
\end{figure}

In this section, we introduce the fundamentals of resilience, including the definition, its relationships to other uncertainty-aware terminologies, enabling mechanisms, and quantification metrics. We start with the concept of uncertainty and the necessity of uncertainty awareness, as illustrated in Fig. \ref{fig:building_diagram}. The key information in this section is summarized in Fig. \ref{fig:key_information_fundamentals}.

\subsection{Uncertainties and Uncertainty-Awareness}\label{sec:uncertainties}
Uncertainties reflect the discrepancies between human knowledge and underlying realities, such as model mismatches, channel disturbances, parameter estimation errors, link blockages, interference, antenna faults, backhaul failures, power outages, cell congestion, jamming, attacks, and network topological variations \cite{wang2025uncertainty}. Some uncertainties are aleatoric (e.g., inherent stochasticity in varying environments), whereas others are epistemic (e.g., limited knowledge due to inadequate data collection); some uncertainties are natural, while others are artificial, regardless of whether they are adversarial, benign, or neutral. These uncertainties widely exist in diverse wireless systems of different scales, ranging from point-to-point physical-layer (PHY) links, to single-station cells, and to cellular, sensor, and non-terrestrial networks (NTNs). If not properly handled, these uncertainties can cause performance perturbations and degradations or even operational interruptions and failures in the system of interest \cite{wang2025uncertainty,IMT-2030,ma2026vision}. To this end, uncertainty awareness in wireless systems design and operation warrants thorough investigation, which specifically involves equipping wireless systems with pivotal capabilities such as resilience, reliability, robustness, adaptability, survivability, and recoverability. Here, as per the research scales and focuses,  the connotation of the word ``system" broadly includes both physical entities (e.g., a PHY link, a cell, or a network) and algorithmic procedures (e.g., a routing strategy, a beamforming scheme, an optimization or statistical method, or a computing paradigm) in wireless engineering; cf., e.g., \cite[p.~3]{jackson2013resilience}. The former corresponds to systems design (e.g., architectures), while the latter corresponds to systems operation (e.g., orchestrations). The functional boundary between the two aspects, however, is not typically sharp because the uncertainty awareness of an algorithmic framework can directly lead to that of a physical system: For example, resilient packet routing and user handover algorithms that can perceive and respond to uncertainties contribute to the resilience of wireless networks.

In different contexts and literature, uncertainties are diversely referred to as disturbances, errors, adversities, faults, attacks, threats, interference, unknowns, among many others. As consequences of uncertainties, disruptions denote deviations of a system from its intended operation (a discrete-valued status) or performance (a continuous-valued indicator), in either a degraded, interrupted, or failed manner, corresponding to degradations, interruptions, and failures \cite{jackson2013resilience,wang2025uncertainty}. Note, however, that the conceptual boundaries between these causes and consequences are usually blurred because in cascaded systems, caused statuses (i.e., disruptions) of the preceding modules serve as the causing factors (e.g., adversities) of the succeeding ones. This relationship also extends to surrounded systems, where environmental disruptions act as adversities.

\subsection{Concepts of Resilience and Reliability}\label{sec:resilience}
According to \cite[p.~2]{jackson2013resilience}, \cite[p.~16]{IMT-2030}, and \cite[p.~8]{ma2026vision}, resilience is the system-level capability to maintain and restore service performance (i.e., continue functioning) in dynamic and uncertain environments throughout the system's lifecycle. Depending on contexts, its specific connotations (or objectives) include preventing the system from encountering uncertainties, adapting the system to changing conditions, maintaining performance during disturbances (e.g., slight uncertainties), restoring performance from disruptions (e.g., after strong uncertainties), among others. In some literature, the narrow-sense definition of resilience, i.e., restorability, due to its dictionary meaning, also applies \cite[p.~1]{jackson2013resilience}, \cite[p.~32]{wang2025uncertainty}. In future wireless engineering, as IMT-2030 does, resilience is suggested to be interpreted in the broad sense by default. Similar to green communications and artificial intelligence (AI)-native networks, resilience emerges as a new design paradigm for future wireless systems.

\begin{figure}[!htbp]
    \centering
    \includegraphics[width=0.4\textwidth]{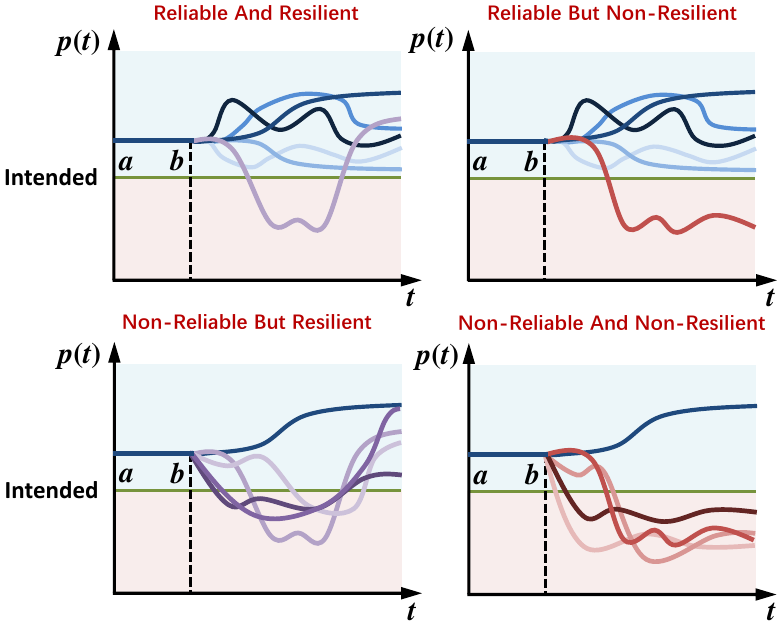}
    \caption{The conceptual comparison of reliability and resilience, leveraging the performance indicator $p(t)$ of the system over time $t$. The uncertainties occur at the point $b$ and cause stochasticity in performance trajectories. Reliability is a statistical metric quantifying the likelihood that the system's performance trajectory will drop below the intended level. Resilience, however, is a trajectory-wise notion, describing whether a trajectory can be maintained above the intended level or restored to above the intended level if the trajectory drops below it.}
    \label{fig:reliability_vs_resilience}
\end{figure}

Reliability is another system-level capability: In general systems engineering, it is defined as the probability that a system maintains its intended function, operation, or performance for a given period of time. 
In wireless communication systems, however, the notion of reliability is often used in a narrow sense at the physical-link level, i.e., ``the capability of transmitting a predefined amount of data successfully within a predetermined time duration with a given probability" \cite[p.~16]{IMT-2030}, \cite[p.~7]{ma2026vision}, which is associated with the usage scenario of ``hyper reliable and low-latency communication (HRLLC)" in IMT-2030. Here, the probability is induced by the randomness of the underlying uncertainties, such as channel disturbances. In future wireless engineering, examining the reliability issues of various systems, beyond packet transmission, is expected. For a conceptual comparison of reliability and resilience, see Fig. \ref{fig:reliability_vs_resilience}. In the upper two panels, most performance trajectories remain above the intended level; thus, the system has a high reliability. In contrast, in the lower two panels, most trajectories drop below the intended level; therefore, the system has a low reliability. In the left two panels, all disrupted trajectories can be restored to above the intended level, indicating that the system is resilient, whereas the system associated with the right two panels is not resilient. Taking packet transmission as an example, the successful probability of $1-10^{-5}$ within the allowed maximum latency can be seen as reliable \cite[p.~7]{ma2026vision}, while equipping with feedback and retransmission mechanisms can be regarded as resilient. Resilience does not imply reliability, and vice versa.

\subsection{Capability Dimensions of Resilience}\label{sec:capability-dimensions}
Resilience includes four key capability dimensions (i.e., particularizations) in different contexts and scenarios: robustness, adaptability, survivability, and recoverability.

Robustness and adaptability represent the system's capabilities to maintain normal operation and intended performance under slight-to-moderate uncertainties. The difference is that robustness enables the system to tolerate or withstand uncertainties without changing its decisions, behaviors, or configurations, whereas adaptability allows the system to adjust its decisions, behaviors, or configurations in response to changing environments or operating conditions \cite[p.~32]{wang2025uncertainty}; see also \cite{wang2026robust} and \cite{wang2025fast}, respectively, for technical discussions. For example, robust people can maintain their health under diverse viral challenges through inherent immune capacities without changing their lifestyle or taking medicines. In contrast, adaptive individuals actively monitor their health conditions and adopt appropriate interventions, such as lifestyle changes or medicines, to respond to evolving viral threats. 

\begin{figure}[!htbp]
    \centering
    \includegraphics[width=0.4\textwidth]{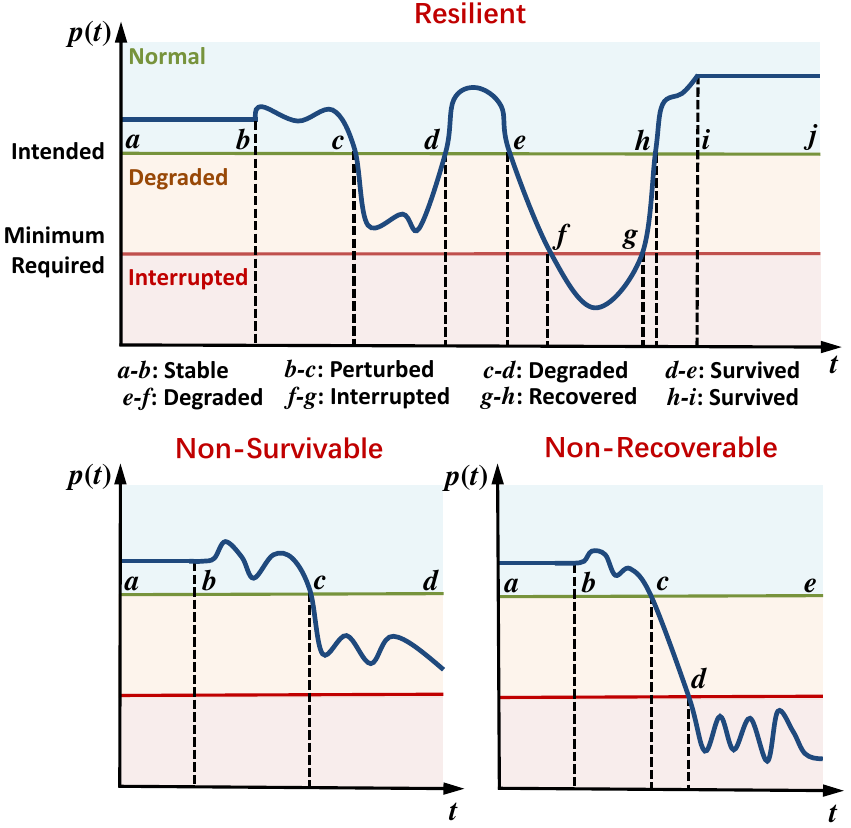}
    \caption{The conceptual comparison of survivability and recoverability. According to the performance indicator $p(t)$ at time $t$, the system has three operating statuses: Normal, Degraded, and Interrupted. In the perturbed phase (e.g., from Point b to Point c), despite uncertainties, the system maintains its intended performance using robust and adaptive strategies. The system associated with the lower two panels is non-resilient because it is either non-survivable or non-recoverable.}
    \label{fig:system_performance}
\end{figure}

When strong uncertainties happen, maintaining normal operation and intended performance may be challenging. In this case, we can instead pursue degraded performance, that is, maintaining the performance above the minimum-required level, rather than the intended level. If the system can restore its performance to above the intended level from the degraded range, it is referred to as survivable. In contrast, if the system can restore its performance to above the intended level from below the minimum-required level, it is referred to as recoverable. For the conceptual comparison, see Fig. \ref{fig:system_performance}. Both survivability and recoverability are called restorability, but they are differentiated in different contexts. Continuing with the previous human health example, restorability reflects a person's recuperation after hospitalization.

Unlike robustness, which primarily supports the retention of operation and performance during disturbances, adaptability can also be a key enabler of restorability after disruptions. A motivating instance is as follows. A robust wireless network may maintain users’ services during a base station (BS) failure through redundant links (e.g., coordinated multi-point transmission), whereas an adaptable network may adjust routing and reallocate resources after the failure to restore the lost coverage and capacity. Hence, adaptation can be conducted before, during, and after an encounter with an uncertainty or disruption. 

\subsection{Supporting Capabilities for Resilience}\label{sec:supporting-capabilities}
To enable adaptability facing uncertainties, sufficient observability of the uncertainties or their effects, such as faults, unknowns, or degraded system states and performance, is essential. Otherwise, systems cannot establish effective feedback or feedforward loops to support informed decisions, behaviors, and configurations \cite[p.~33]{wang2025uncertainty}. In addition to observability, predictability is another system-level perceiving capability to enable proactive protection and adaptation, that is, to forecast upcoming uncertainties (e.g., demands and workloads) and prepare for them in advance. 

For smarter functions, operations, and responses in the future, systems are anticipated to learn from uncertainties and histories. This capability, called evolvability, particularly includes more accurate prediction and observability, more efficient fault-tolerance, adaptation (e.g., scaling and reconfiguration), and faster survival
and recovery. In modern practice, evolvability of systems is largely supported by AI- and machine learning (ML)-driven approaches, although specific evolving strategies in different contexts remain open to be explored.

\subsection{Enabling Mechanisms of Resilience}\label{sec:mechanisms}
As discussed earlier, achieving resilience in a system requires equipping the system with a combination of particularized capabilities, including robustness, adaptability, survivability, and recoverability, which are to be activated under different scenarios and conditions. Moreover, supporting capabilities, such as observability, predictability, and evolvability, are also essential for realizing resilience. In this subsection, we discuss typical mechanisms that enable these capabilities. Here, mechanisms broadly refer to diverse principles of systems design and operation: that is, architecture, module, and component designing, as well as their operational management. Note, however, that the relationships between mechanisms and capabilities are not one-to-one but many-to-many: To enable one capability may involve a batch of mechanisms, and one mechanism can be utilized for assisting multiple capabilities. 

By definition, \textit{tolerance} (also known as \textit{withstanding}) of uncertainties, \textit{adaptation} to changing conditions, \textit{survival} from degradations, and \textit{recovery} from interruptions and failures are naturally the enabling mechanisms of resilience, corresponding to robustness, adaptability, survivability, and recoverability, respectively. We refer to these four mechanisms as objective mechanisms, whereas any other mechanisms that facilitate or realize them are called means mechanisms. In this sense, \textit{prediction} and \textit{observation} fall into means mechanisms. Below, we investigate other typical means mechanisms in general systems engineering that are also implementable in wireless systems; for extensive reading, see, e.g., \cite[pp.~33-34]{wang2025uncertainty}, \cite[Section~6]{jackson2013resilience}, \cite[Section~1]{brtis2016think}. However, the list is only representative and not necessarily exhaustive. Practitioners should identify additional mechanisms in their applications.

\textit{Redundancy}: Providing more than one way to perform a required function, such as backup components or parallel resources. Redundancy can mitigate single-point failures by maintaining functionality when one element fails. In wireless engineering, coordinated multi-point (CoMP) transmission is guided by the redundancy mechanism.

\textit{Diversity}: Achieving redundancy through heterogeneous ways. Diversity can mitigate common-mode failures and shared vulnerabilities that cannot be addressed by homogeneous redundancy. In wireless engineering, interleaving after channel coding and multi-antenna transmission are guided by the diversity mechanism.

\textit{Margin}: Providing more capacity than expected to buffer adverse impacts from uncertainties and mitigate performance degradation. Margin is also known as \textit{buffering} or \textit{absorption}, which enhances robustness against uncertainties, such as aging effects. In wireless engineering, inserting guard times or guard bands is guided by the margin mechanism.

\textit{Modularization}: Disaggregating and segmenting a system into loosely coupled functional elements such that changes or failures in one element have limited impacts on others. Modularization establishes functional boundaries, reduces damage propagation, and facilitates fault localization and isolation. For example, conventional modularized wireless architectures built upon functional blocks can be more maintainable and resilient than tightly coupled (i.e., monolithic or end-to-end) AI architectures.

\textit{Decentralization}: Distributing system functionality across multiple nodes, locations, or entities such that the loss or degradation of individual elements does not compromise overall operation. Decentralization reduces dependence on single nodes and improves resilience through distributed communication, cooperation, and decision-making. In this sense, mesh networks can be more resilient to access-point (AP) failures than cellular networks.

\textit{Isolation}: Limiting the propagation of failures, faults, or attacks among system components through physical or logical separation. Isolation, also called \textit{separation}, reduces harmful interactions among elements that have different protection requirements. Modularized and decentralized system architectures inherently support the isolation mechanism, justifying their widespread deployment in wireless practice. In wireless engineering, beamforming provides spatial isolation by steering beams toward intended users and nulls toward interferers.

\textit{Countermeasure}: Applying defensive actions to confuse, deceive, or impede adversaries and reduce the effectiveness of intentional threats. Countermeasure may also be known as \textit{suppression} or \textit{compensation} in different contexts. In wireless engineering, anti-jamming null steering and interference cancellation are guided by the countermeasure mechanism. 

\textit{Estimation}: Acquiring information about unknown system states, parameters, or uncertainties from measurements. It can also be called \textit{detection}, \textit{testing}, or \textit{monitoring}, among many others, in different technical contexts. In wireless communications, channel estimation, interference estimation, and state estimation are representative examples. Estimation is a particularization of the observation mechanism (which can also broadly include anomaly detection and fault diagnosis) and serves as an enabling mechanism for resilience, e.g., by facilitating adaptation.

\textit{Avoidance}: Preventing or reducing exposure to predicted disturbances, threats, or adverse conditions by proactively changing system operation, configuration, or behavior. In wireless communications, for instance, a mobile user can walk around large buildings to avoid link blockages. Avoidance is a proactive particularization of adaptation before adverse impacts happen.

\textit{Scaling}: Accommodating changes in system size, workload, or resource demand (e.g., burst joining and exiting of users). Scaling is a reactive particularization of adaptation. In wireless communications, a scalable base station can support varying user densities and loads.

\textit{Reconfiguration}: Modifying the system's configuration, e.g., network topologies, operating modes, input parameters, functional relationships, and organizational structures. Reconfiguration is also a reactive particularization of adaptation and may be referred to as \textit{restructuring}, \textit{realignment}, \textit{reallocation}, or \textit{reorganization} in different contexts. The conceptual boundary between scaling and reconfiguration is not sharp: Reconfiguration can serve as one of the enabling mechanisms of scalability, for example, through resource scheduling in decentralized networks as new nodes join.

\textit{Replacement}: Substituting failed, degraded, or unavailable elements with new ones to restore expected operation and performance. Replacement is also a reactive particularization of adaptation, but it happens after system interruptions or failures. In wireless systems, vehicle-mounted BSs can temporarily replace damaged terrestrial ones after disruptions due to, e.g., sabotage and natural disasters. 


\textit{Coordination}: Integrating multiple complementary mechanisms across modules, layers, and algorithms to achieve resilience. For example, a resilient wireless network always coordinates physical-layer adaptation, access-layer routing decisions, and application-layer resource management to mitigate the impact of dynamic disruptions.

Note again that these generic mechanisms apply to a wide range of systems at different scales, including both physical entities and algorithmic procedures. Taking the modularization mechanism as an example, both a virtual software system and a physical communication system can be modularized into interconnected functional blocks to improve resilience.

\subsection{Quantification of Resilience}\label{sec:quantification}
\begin{figure}[!htbp]
    \centering    \includegraphics[width=0.4\textwidth]{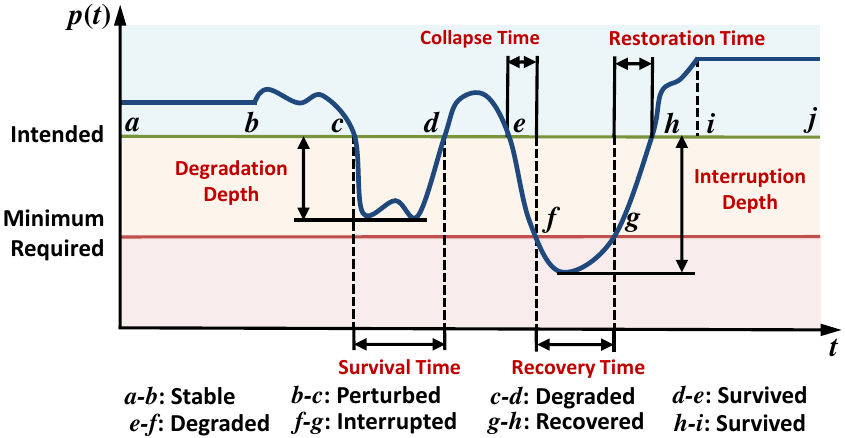}
    \caption{Representative metrics of resilience, including Degradation Depth, Interruption Depth, Survival Time, Collapse Time, Recovery Time, and Restoration Time. Except for collapse time, the smaller the values of the other quantities, the more resilient the system is.}
    \label{fig:resilience_metrics}
\end{figure}

The key quantification metrics of resilience include degradation depth, interruption depth, survival time, collapse time, recovery time, and restoration time, wherever they are applicable; see Fig. \ref{fig:resilience_metrics}. For better resilience, the collapse time should be as large as possible, while the other values should be as small as possible. 

\section{Resilience of Specific Wireless Systems}\label{subsec:connotations}
Depending on the research and development focus, resilience can be realized at different scales of wireless systems, including PHY links, single-station systems, terrestrial (e.g., cellular, cell-free, and sensor) networks, and non-terrestrial networks (e.g., unmanned aerial vehicle (UAV)- and satellite-involved), by particularizing the previously discussed objective and means mechanisms in different technical contexts. Note that resilience can also be realized in the algorithmic procedures that underpin these physical entities. In this section, as illustrating examples, we discuss the specific connotations and enabling techniques of resilience in two representative types of wireless systems: PHY links and UAV networks. Here, ``techniques" refer to the concrete solutions that realize, implement, or particularize the generic objective and means mechanisms of resilience in specific systems and technical contexts.

\subsection{Resilience of PHY Links}
A PHY link comprises a transmitter, a radio propagation channel, a receiver, and the associated signal-processing procedures required to convey information between the transmitter and receiver.

Sources of uncertainty at the PHY-link level include imperfect or outdated channel state information (CSI) due to channel estimation errors or pilot contamination and deep or rapidly varying fading, sudden link blockage or path birth and death, excessive interference or intentional jamming, mobility-induced Doppler shifts, synchronization errors, oscillator phase noise, carrier-frequency offset, and radio-frequency (RF) hardware impairments. 

The link status or utility is typically assessed using service-oriented metrics such as block error rate, effective throughput (goodput), latency, outage probability and duration, and, when confidentiality is required, secrecy rate or secrecy outage probability. Signal-to-noise ratio, signal-to-interference-plus-noise ratio (SINR), and achievable rate provide useful and analytically tractable PHY-layer indicators, although they alone do not determine whether the required communication service remains available.

Resilience in this context refers to the capability of a PHY link to sustain a required or minimally acceptable level of information delivery under uncertainties, adapt its transmission and reception configurations (e.g., time, frequency, space, power resources) as conditions change, and restore decodable communication after an interruption. At the PHY-link level, controllable degrees of freedom include waveform, pilot design, channel coding, beamforming, transmit power, bandwidth, blocklength, carrier frequency, modulation and coding scheme (MCS), CSI acquisition and tracking, receiver processing, and propagation configuration. These degrees of freedom provide the entry points for implementing concrete resilience mechanisms and techniques. To be specific, {robustness} limits immediate performance degradation caused by channel uncertainty and RF impairments through channel coding (e.g., low-density parity-check (LDPC) code, polar code) and equalization, uncertainty-aware or quality-of-service (QoS)/SINR-constrained beamforming, robust waveform design, robust power or spectrum allocation, and diversity transmission (e.g., space-time block coding (STBC)). {Adaptability} enables the transmitter and receiver to adjust MCS, blocklength, transmit power, bandwidth, beam direction, beam-tracking policy, receiver combining and equalization strategies, or propagation configuration as channel conditions and service requirements change. {Survivability} preserves a minimum information-delivery capability during severe degradation and can be enhanced through MCS fallback, packet repetition, or hybrid automatic repeat request (HARQ). {Recoverability} restores a decodable and synchronized link after an outage through channel reacquisition or re-estimation, beam-failure recovery, resynchronization, initial access or link re-establishment, carrier switching, retransmission, or the activation of alternative propagation paths assisted by either active devices (e.g., relays) or passive ones (e.g., reconfigurable intelligent surfaces (RISs), beyond-diagonal RIS (BD-RIS), and backscatter devices). 

For concrete techniques that correspond to the means mechanisms for PHY links, see Table \ref{tab:phy_ntn}.

\begin{table}[!htbp]
    \centering
    \caption{Specific techniques that correspond to means mechanisms in PHY links
    and UAV networks}
    \label{tab:phy_ntn}
    \scriptsize
    \setlength{\tabcolsep}{2.5pt}
    \renewcommand{\arraystretch}{1.15}

    \setlength{\arrayrulewidth}{0.4pt}

    \begin{tabular}{
        >{\centering\arraybackslash}m{0.25\columnwidth}
        |
        >{\centering\arraybackslash}m{0.29\columnwidth}
        |
        >{\centering\arraybackslash}m{0.29\columnwidth}
    }
        \thickhline

        \multicolumn{1}{c|}{\multirow{2}{*}{\textbf{Means Mechanism}}} & \multicolumn{2}{c}{\textbf{Specific Techniques}}                         \\ \cline{2-3}
        \multicolumn{1}{c|}{}                   & \multicolumn{1}{c|}{\textbf{PHY Links}} & \multicolumn{1}{c}{\textbf{UAV Networks}} \\ \hline

        Prediction
        & Blockage prediction
        & Traffic-demand/coverage-hole prediction\\ \hline

        Observation \& Estimation
        & Channel estimation, MIMO detection,  beam tracking
        & Link-visibility, node health monitoring, localization \\ \hline

        Redundancy
        & LDPC/polar coding, HARQ, preamble/control repetition, antenna selection
        & Backup UAVs \\ \hline

        Diversity
        & STBC, interleaving
        & Heterogeneous air-ground access/backhaul paths\\ \hline

        Margin
        & Power/beam/blocklength allocation, conservative MCS
        & Battery and coverage-overlap reserves \\ \hline

        Modularization
        & Group-connected BD-RIS
        & Cluster-based swarm organization \\ \hline

        Decentralization
        & N/A
        & Leaderless distributed UAV topology control \\ \hline

        Isolation
        & N/A
        &Compromised UAV quarantine, control–data plane isolation \\ \hline

        Countermeasure
        & Anti-jamming/null-steering beam design, Doppler-resilient modulation
        & Cooperative anti-jamming UAV deployment \\ \hline

        Avoidance
        & Interference-aware channel/beam selection
        & Hazard/obstacle-aware trajectory planning,  no-fly-zone avoidance\\ \hline

        Scaling
        & Limited feedback, resource-block allocation, hybrid beamforming 
        &  Autonomous on-demand fleet deployment\\ \hline

        Reconfiguration
        & MCS fallback, RIS, bandwidth/blocklength adjustment 
        &  Failure-triggered topology/route reconfiguration \\ \hline

        Replacement
        & Carrier/beam/propagation switching 
        &  Standby-UAV takeover\\ \hline

        Coordination
        & Synchronization
        &  Joint multi-UAV placement and resource scheduling\\
        \thickhline
    \end{tabular}
\end{table}

\subsection{Resilience of UAV Networks}
UAV networks represent a typical example of NTNs. A UAV network comprises multiple aerial nodes operating as airborne BSs, relays, wireless-fronthaul nodes, or edge-computing platforms interconnected through air-to-air and air-to-ground links.

UAV networks operate under uncertainties arising from time-varying communication environments and sensing workloads, dynamic topology, and imperfect knowledge of position, attitude, and mobility. Their air-to-air and air-to-ground links may also be disrupted by extreme weather, obstacles, and intermittent connectivity. UAV-specific challenges further include uncertain battery levels and component health, incomplete or delayed local information, and security threats such as jamming, global navigation satellite system spoofing, malicious control messages, and compromised nodes.

Utilities for UAV networks should capture network-wide service continuity, mission effectiveness, and operational sustainability, using metrics including mission-completion probability, end-to-end latency,  service outage probability, and network lifetime. Coverage, graph connectivity, backhaul capacity, UAV availability, and energy consumption are informative network-state indicators, although they do not individually establish service continuity.

UAV-network resilience is the ability to sustain essential coordination functions (e.g., communication, sensing, computing), adapt aerial resources (e.g., node availability, trajectories, onboard energy) to changing conditions, survive severe disruptions, and restore lost service or mission capabilities. Unlike PHY-link resilience, it exploits network-level degrees of freedom in node association, inter-UAV coordination, aerial fronthaul, trajectories, topology, relay roles, and task allocation. These degrees of freedom support four capability dimensions through specific resilience mechanisms and enabling techniques. Robustness limits initial degradation through backup UAVs, redundant topologies, diverse fronthaul paths, energy and coverage margins, and risk- or no-fly-zone-aware trajectory planning. Adaptability responds to evolving conditions and threats through state estimation, traffic prediction, resource reallocation, fronthaul switching, and cooperative anti-jamming. Survivability maintains minimum critical services after severe node or link losses through decentralized operation, failure-triggered topology reconfiguration, compromised-node isolation, priority-aware task allocation, and multi-hop relaying. Recoverability restores the target service through standby-UAV takeover, UAV replacement or redeployment, route and fronthaul restoration, and reassignment of users, relay roles, and tasks.

For concrete techniques that correspond to the means mechanisms for UAV networks, see Table \ref{tab:phy_ntn}.

\section{Trade-Offs in Resilient Wireless Systems}
Although resilience is a fundamental capability requirement for future wireless systems, it is not achieved without cost: Improving resilience requires balancing it against other desirable system objectives or properties. Some representative examples are as follows. 
\begin{itemize}
    \item \textit{Resource Overhead}: Achieving resilience incurs resource overhead since mechanisms such as redundancy, diversity, and margin consume additional spectrum, energy, hardware, communication, and computational resources. For example, resilience of wireless systems entails information-acquisition overhead because observing, estimating, or predicting unknown system states, parameters, and environmental conditions requires additional pilots, feedback, signaling, and computation (e.g., computing times). 

    \item \textit{Nominal Performance Sacrifice}: Resilience may compromise nominal performance. For example, robust solutions are inherently conservative, which may underperform nominal solutions when anticipated uncertainties do not materialize as they are random. Note that robust solutions are usually optimized for worst-case conditions, whereas nominal solutions are for nominal conditions. For another example, modular architectures can facilitate fault localization, isolation, replacement, maintenance, and reconfiguration, but interfaces between modules can constrain joint performance optimization. Hence, highly integrated or end-to-end designs may achieve better nominal performance, whereas modular designs can offer better maintainability and resilience.

    \item \textit{Complexity Growth}: Resilience often increases implementation complexity and coordination difficulty, as reconfiguration, heterogeneous operation, and distributed cooperation require more complex decision-making, information exchange, and synchronization. For example, decentralization removes single points of failure and distributes functionality, but distributed nodes must exchange information and coordinate decisions. This introduces signaling overhead, synchronization issues, consensus latency, and potentially suboptimal decisions compared with perfectly informed centralized control.
    
    \item \textit{Unprofitable Intervention}:
    Proactive resilience introduces a trade-off between timely intervention and unnecessary intervention: Acting sufficiently early can prevent or mitigate anticipated disruptions, whereas actions triggered by inaccurate predictions or false alarms may waste resources or even degrade nominal performance.

    \item \textit{Delayed Response}: Resilience requires prompt actions after a disruption. However, waiting for more measurements or performing more sophisticated optimization can improve decision quality. Hence, systems face a trade-off between reaction latency and solution quality. 

    \item \textit{Global Consideration}: A node or module acting to protect itself can shift the burden elsewhere. For example, increasing transmit power can restore one link but increase interference to neighboring users; rerouting traffic around one congested link can overload another. Therefore, resilience actions should be considered at the global level rather than only locally.
\end{itemize}
As indicated, resilience more often than not trades off against resource efficiency, information acquisition, nominal performance, implementation and coordination complexity, latency, energy consumption, and risks arising from unnecessary or overly conservative resilience actions. Therefore, resilience engineering is fundamentally a problem of managing these competing objectives, rather than pursuing resilience alone.

\section{Conclusions}\label{sec:conclusion}
This paper presents a systematic framework for resilience in wireless systems from a trustworthiness perspective. We clarified the concept of resilience, established a hierarchical framework that organizes its capability dimensions, enabling mechanisms, and implementation techniques, identified its key quantification metrics, illustrated its realization in representative wireless systems, and discussed its associated engineering trade-offs. Through the proposed framework, we aim to provide a useful conceptual foundation for the analysis, design, and evaluation of resilient wireless systems.

Despite this progress, resilience remains at an early stage of research and development in wireless engineering. Future study should focus on establishing rigorous mathematical foundations for resilience, including formal definitions, performance metrics, and analytical models that quantitatively characterize resilience under specific scenarios and diverse uncertainties. Another important direction is the development of resilience-oriented statistical, optimization, learning, and control algorithms that jointly optimize resilience with competing objectives such as resource efficiency, latency, and implementation complexity. Finally, standardized resilience benchmarks, evaluation methodologies, and practical design principles are needed to enable fair comparison and systematic deployment of resilient wireless systems in future networks.

\bibliographystyle{IEEEtran}
\bibliography{References}

@article{guo2020explainable,
  title={Explainable artificial intelligence for {6G}: Improving trust between human and machine},
  author={Guo, Weisi},
  journal={IEEE Commun. Mag.},
  volume={58},
  number={6},
  pages={39--45},
  year={2020},
  publisher={IEEE}
}

@misc{IMT-2030,
  author       = {{International Telecommunication Union (ITU)}},
  title        = {Framework and overall objectives of the future development of {IMT} for 2030 and beyond},
  year         = {2023},
  howpublished = {\url{https://www.itu.int/rec/R-REC-M.2160/en}},
  note         = {Accessed: 2026-07-09}
}

@article{ma2026vision,
  title={From Vision to Design Targets: Technical Performance Requirements for {IMT-2030}},
  author={Ma, Liang and Grant, Marc and Lin, Hui and Sk{\"o}ld, Johan and Liu, Ruiqi and Shao, Jiafeng},
  journal={IEEE Commun. Mag.},
  volume={64},
  number={6},
  pages={6--9},
  year={2026},
  publisher={IEEE}
}

@ARTICLE{wang2025uncertainty,
  author={Wang, Shixiong and Dai, Wei and Sun, Jianyong and Xu, Zongben and Li, Geoffrey Ye},
  journal={IEEE Commun. Mag.}, 
  title={Uncertainty Awareness in Wireless Communications and Sensing}, 
  year={2026},
  volume={64},
  number={2},
  pages={30-38},
  doi={10.1109/MCOM.001.2400714}}

@article{huaizhou2013fairness,
  title={Fairness in wireless networks: Issues, measures and challenges},
  author={Huaizhou, SHI and Prasad, R Venkatesha and Onur, Ertan and Niemegeers, IGMM},
  journal={IEEE Commun. Surveys Tuts.},
  volume={16},
  number={1},
  pages={5--24},
  year={2013},
  publisher={IEEE}
}

@article{chen2011fundamental,
  title={Fundamental trade-offs on green wireless networks},
  author={Chen, Yan and Zhang, Shunqing and Xu, Shugong and Li, Geoffrey Ye},
  journal={IEEE Commun. Mag.},
  volume={49},
  number={6},
  pages={30--37},
  year={2011},
  publisher={IEEE}
}

@ARTICLE{wang2025fast,
  author={Wang, Ouya and He, Hengtao and Zhou, Shenglong and Ding, Zhi and Jin, Shi and Letaief, Khaled B. and Li, Geoffrey Ye},
  journal={IEEE Commun. Mag.}, 
  title={Fast Adaptation for Deep Learning-Based Wireless Communications}, 
  year={2025},
  volume={63},
  number={10},
  pages={158-164},
  doi={10.1109/MCOM.001.2400502}}

@article{zou2016survey,
  title={A survey on wireless security: Technical challenges, recent advances, and future trends},
  author={Zou, Yulong and Zhu, Jia and Wang, Xianbin and Hanzo, Lajos},
  journal={Proc. IEEE},
  volume={104},
  number={9},
  pages={1727--1765},
  year={2016},
  publisher={IEEE}
}

@article{qu2018privacy,
  title={Privacy of things: Emerging challenges and opportunities in wireless internet of things},
  author={Qu, Youyang and Yu, Shui and Zhou, Wanlei and Peng, Sancheng and Wang, Guojun and Xiao, Ke},
  journal={IEEE Wireless Commun.},
  volume={25},
  number={6},
  pages={91--97},
  year={2018},
  publisher={IEEE}
}

@article{thomas2024causal,
  title={Causal reasoning: Charting a revolutionary course for next-generation AI-native wireless networks},
  author={Thomas, Christo Kurisummoottil and Chaccour, Christina and Saad, Walid and Debbah, Merouane and Hong, Choong Seon},
  journal={IEEE Veh. Technol. Mag.},
  volume={19},
  number={1},
  pages={16--31},
  year={2024},
  publisher={IEEE}
}

@article{xiao2008accountability,
  title={Accountability for wireless {LANs}, ad hoc networks, and wireless mesh networks},
  author={Xiao, Yang},
  journal={IEEE Commun. Mag.},
  volume={46},
  number={4},
  pages={116--126},
  year={2008},
  publisher={IEEE}
}

@article{rahmani2023next,
  title={Next-generation {IoT} devices: Sustainable eco-friendly manufacturing, energy harvesting, and wireless connectivity},
  author={Rahmani, Hamed and Shetty, Darshan and Wagih, Mahmoud and Ghasempour, Yasaman and Palazzi, Valentina and Carvalho, Nuno B and Correia, Ricardo and Costanzo, Alessandra and Vital, Dieff and Alimenti, Federico and others},
  journal={IEEE J. Microw.},
  volume={3},
  number={1},
  pages={237--255},
  year={2023},
  publisher={IEEE}
}

@article{jackson2013resilience,
  title={Resilience principles for engineered systems},
  author={Jackson, Scott and Ferris, Timothy LJ},
  journal={Systems Engineering},
  volume={16},
  number={2},
  pages={152--164},
  year={2013},
  publisher={Wiley Online Library}
}

@article{wang2026robust,
  title={Robust Processing and Learning: Principles, Methods, and Wireless Applications},
  author={Wang, Shixiong and Dai, Wei and Wang, Li-Chun and Li, Geoffrey Ye},
  journal={arXiv preprint arXiv:2602.09848},
  year={2026}
}

@techreport{brtis2016think,
  author       = {John S. Brtis},
  title        = {How to Think About Resilience in a {DoD} Context: A {MITRE} Recommendation},
  institution  = {The MITRE Corporation},
  address      = {Colorado Springs, CO, USA},
  type         = {Technical Report},
  number       = {MTR-160138},
  year         = {2016}
}




\end{document}